\documentclass[twocolumn]{aastex6}

\fullcollaborationName{The Friends of AASTeX Collaboration}

\begin{document}


\title{Far-infrared properties of infrared bright dust-obscured galaxies selected with IRAS and AKARI far-infrared all-sky survey}



\author{Yoshiki Toba 		\altaffilmark{1},
		Tohru Nagao 		\altaffilmark{2},
		Wei-Hao Wang		\altaffilmark{1},
		Hideo Matsuhara 	\altaffilmark{3,4},		
   		Masayuki Akiyama	\altaffilmark{5},
   		Tomotsugu Goto	    \altaffilmark{6},		
   		Yusei Koyama	    \altaffilmark{7,8},
   		Youich Ohyama	    \altaffilmark{1},			
   		Issei Yamamura  	\altaffilmark{3,4}
	 	}	  
\affil{}  			  
  \altaffiltext{1}{Academia Sinica Institute of Astronomy and Astrophysics, PO Box 23-141, Taipei 10617, Taiwan}
  \email{toba@asiaa.sinica.edu.tw}
  \altaffiltext{2}{Research Center for Space and Cosmic Evolution, Ehime University, Bunkyo-cho, Matsuyama, Ehime 790-8577, Japan} 
  \altaffiltext{3}{Institute of Space and Astronautical Science, Japan Aerospace Exploration Agency, 3-1-1 Yoshinodai, Chuo-ku, Sagamihara, Kanagawa 252-5210, Japan}
  \altaffiltext{4}{Department of Space and Astronautical Science, SOKENDAI (The Graduate University for Advanced Studies), 3-1-1 Yoshinodai, Chuo-ku, Sagamihara, Kanagawa 252-5210, Japan}
  \altaffiltext{5}{Astronomical Institute, Tohoku University, Aramaki, Aoba-ku, Sendai 980-8578, Japan}
  \altaffiltext{6}{Institute of Astronomy and Department of Physics, National Tsing Hua University, No. 101, Section 2, Kuang-Fu Road, Hsinchu 30013, Taiwan}    
  \altaffiltext{7}{Subaru Telescope, National Astronomical Observatory of Japan, National Institutes of Natural Sciences, 650 North A'ohoku Place, Hilo, HI 96720, U.S.A. }
  \altaffiltext{8}{Graduate University for Advanced Studies (SOKENDAI), Osawa 2-21-1, Mitaka, Tokyo 181-8588, Japan} 









\begin{abstract}

We investigate the star forming activity of a sample of infrared (IR)-bright dust-obscured galaxies (DOGs) that show an extreme red color in the optical and IR regime, $(i - [22])_{\rm AB} > 7.0$.
Combining an IR-bright DOG sample with the flux at 22 $\micron$ $>$ 3.8 mJy discovered by \cite{Toba_16} with {\it IRAS} faint source catalog version 2 and {\it AKARI} far-IR (FIR) all-sky survey bright source catalog version 2, we selected 109 DOGs with FIR data. 
For a subsample of 7 IR-bright DOGs with spectroscopic redshift ($0.07 < z < 1.0$) that was obtained from literature, we estimated their IR luminosity, star formation rate (SFR), and stellar mass based on the spectral energy distribution fitting.
We found that (i) {\it WISE} 22 $\micron$ luminosity at observed frame is a good indicator of IR luminosity for IR-bright DOGs and (ii) the contribution of active galactic nucleus (AGN) to IR luminosity increases with IR luminosity.
By comparing the stellar mass and SFR relation for our DOG sample and literature, we found that most of IR-bright DOGs lie significantly above the main sequence of star-forming galaxies at similar redshift, indicating that the majority of {\it IRAS}- and/or {\it AKARI}-detected IR-bright DOGs are starburst galaxies. 
\end{abstract}




\section{Introduction}
The stellar mass ($M_*$) and the star formation rate (SFR) are two of the most fundamental and important physical quantities of galaxies.
Since a tight correlation between $M_*$ and SFR of galaxies has been discovered \citep[e.g, ][]{Brinchmann}, many authors have intensively investigated this relation for various galaxies at various redshift \citep[e.g, ][]{Daddi,Elbaz,Noeske}.
It is well known that the majority of galaxies follows a relation so-called ``main sequence (MS)'' and this correlation is seen to evolve towards high redshift across all environments \citep[e.g., ][]{Whitaker,Koyama_13,Lee_15,Tomczak}.
However, a comprehensive implication of the tight correlation between stellar mass and SFR and of its redshift evolution is still unclear \citep[see ][]{Casey}.
In addition, it is known that galaxies undergoing active star formation (SF) that could be induced by major merger process lie significantly above the MS and referred as starburst galaxies.
Investigating the relation of these starburst galaxies and MS is important to understand the origin of the $M_*$--SFR connection. 

In this work, we focus on dust-obscured galaxies \citep[DOGs: ][]{Dey}.
Their mid-infrared (MIR) flux densities are three orders of magnitude larger than those at optical wavelengths, implying that a significant active galactic nucleus (AGN) and/or SF activities heat dust.
The optical and ultraviolet (UV) emission originated from these activities is absorbed by heavily surrounding dust that re-emits in the IR wavelength.
Their IR luminosity often exceeds $10^{12}$ $L_{\sun}$ that is classified as ultraluminous IR galaxies \citep[ULIRGs:][]{Sanders}. 
Recently, \cite{Riguccini} investigated the far-IR (FIR) properties for a sample of 95 DOGs within the COSMOS field,  based on spectral energy distribution (SED) fitting.
However, their DOG sample is limited to those with flux density at 24 $\micron$ less than 3.0 mJy (its mean value is $\sim$ 0.4 mJy).
On the other hand, IR-bright DOGs with a much higher MIR flux density are thought to be a maximum phase of SF and AGN activity \citep[e.g., ][]{Hopkins}, ant thus they are likely to be a crucial population to understand what kinds of physical processes drive the SFR--$M_*$ relation.
Recently, we successfully discovered a large number of IR-bright DOGs and investigated their statistical properties \citep{Toba,Toba_16,Toba_17}.
However, their SF properties are still unknown because we lack deep and wide FIR data that are responsible for the SF activity.

In order to estimate the FIR luminosity of IR-bright DOGs and investigate their SF properties, we utilized data from the {\it Infrared Astronomical Satellite} ({\it IRAS}) and {\it AKARI} satellite.
{\it IRAS} is the first satellite that performed an all-sky survey in four IR bands centered at 12, 25, 60, and 100 $\micron$ \citep{Neugebauer,Beichman}.
In this work, we utilized the IRAS Faint Source Catalogue (FSC), version 2.0 \citep{Moshir} reaching a depth of $\sim$0.2 Jy at 12, 25 and 60 $\micron$ and $\sim$1.0 Jy at 100 $\micron$. 
{\it AKARI} is the first Japanese space satellite dedicated to IR astronomy, that was launched in 2006 \citep{Murakami}.
{\it AKARI} performed an all-sky survey at 9, 18, 65, 90, 140, and 160 $\micron$ whose spatial resolution and sensitivity are much higher than those of the {\it IRAS}.
In this work, we utilized the {\it AKARI} Far-Infrared Surveyor (FIS: \citealt{Kawada}) bright source catalogue (BSC) version 2.0 (Yamamura et al. in prep.), which provides the positions and flux densities in the four FIR wavelengths centered at 65, 90, 140, and 160 $\micron$.
The 5$\sigma$ sensitivity at each band is about 2.4, 0.55, 1.4, and 6.3 Jy, which is the deepest data in terms of the FIR all-sky data, and thus these data should be useful to derive the total IR luminosity and SFR of IR-bright DOGs.

In this paper, we present the IR luminosity, stellar mass, and SFR for IR-bright DOGs detected by the Sloan Digital Sky Survey \citep[SDSS: ][]{York}, {\it Wide-field Infrared Survey Explorer} \citep[{\it WISE}: ][]{Wright}, and at least detected by {\it IRAS} or {\it AKARI} FIR all-sky survey.
These multi-wavelength data  are critical for investigating where IR-bright DOGs lie in the SFR--stellar mass (SFR--$M_*$) plane.
Throughout this paper, we adopt $H_0$ = 70 km s$^{-1}$ Mpc$^{-1}$, $\Omega_M$ = 0.3, and $\Omega_{\Lambda}$ = 0.7. 
Unless otherwise noted, all magnitudes refer on the AB system.

\section{Data and analysis}

\subsection{Sample selection}
\label{Selection}
\begin{figure*}
   \centering
   \includegraphics[width=0.8\textwidth]{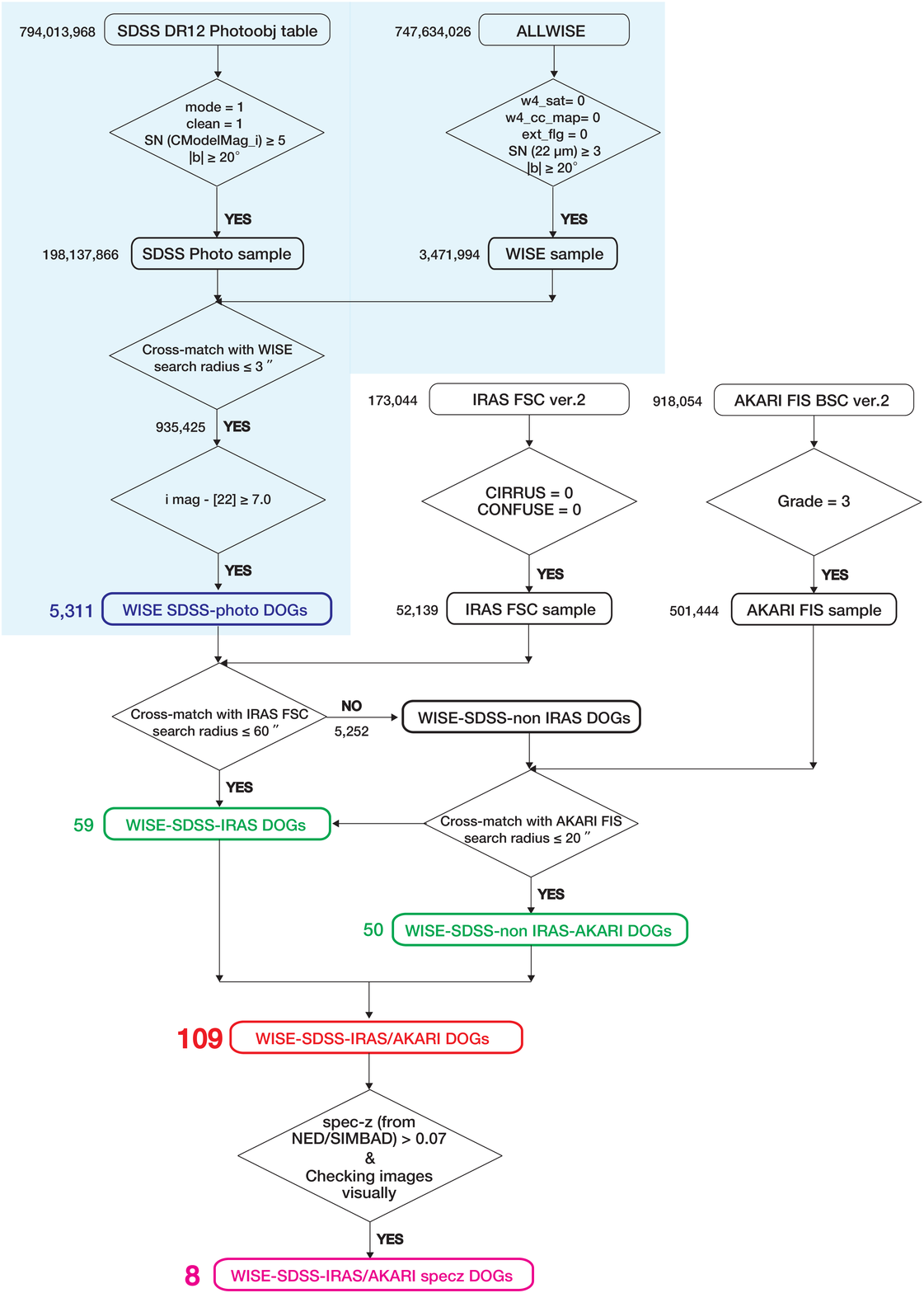}
   \caption{Flow chart of our DOGs selection process. Numbers in this figure denote the number of selected objects at each step. The blue-shaded part is exactly the same as Figure 1 of \cite{Toba_16}.}
   \label{Sample}
   \end{figure*}

We selected 8 IR-bright DOGs with spectroscopic information based on {\it WISE}, SDSS, {\it IRAS}, and {\it AKARI} catalogs \footnote{For the selection process, we employed the TOPCAT, which is an interactive graphical viewer and editor for tabular data \citep{Taylor}.}.
The flow chart of our sample selection process is shown in Figure \ref{Sample}.

The DOG parent sample (hereafter WISE--SDSS photo DOGs) was selected from \cite{Toba_16} who discovered 5,311 IR-bright DOGs with $i - [22] > 7.0$ and flux at 22 $\micron$ $>$ 3.8 mJy, where $i$ and [22] are $i$-band and 22 $\micron$ AB magnitudes, respectively, based on the ALLWISE \citep{Cutri} and SDSS Data Release 12 \citep[SDSS DR12:][]{Alam} catalogs.
For them, we fist cross-identified {\it IRAS} FSC version 2 that includes 173,044 sources.
Before cross-matching, we conservatively selected 52,139 sources that are not affected by cirrus and confusion by adopting {\tt CIRRUS} = 0 and {\tt CONFUSE} = 0.
Using a matching radius of 1$\arcmin$, 59 DOGs (hereafter WISE--SDSS-IRAS DOGs) were selected.
Note that we checked the quality of {\it IRAS} flux in each band ({\tt fqual\_12/25/60/100}) for them, and confirmed that flux at least in one band is measured with good quality (i.e., {\tt fqual\_12/25/60/100} $\geq$ 2).
For $5,311 - 59 = 5,252$ DOGs that are not cross-identified with {\it IRAS} FSC (hereafter WISE--SDSS-non IRAS DOGs), we cross-identified {\it AKARI} FIR BSC version 2 that includes 918,054 sources.
Before cross-matching, we limited ourselves to 501,444 sources with high detection reliability ({\tt GRADE} = 3), i.e., detected by at least two wavelength bands or in four or more scans in one wavelength band. 
Using a matching radius of 20$\arcsec$ which is determined by considering the point spread function size of $\sim$40$\arcsec$ of the {\it AKARI} 90 $\micron$ data, 50 DOGs (hereafter WISE--SDSS-non IRAS--AKARI DOGs) were selected.
Only one {\it AKARI} object has 2 counterpart candidates of WISE-SDSS photo DOGs within the search radius.
We choose the nearest one as a counterpart.
Note that we also cross-identified with {\it AKARI} FSC BSC ver.2. even for WISE--SDSS-IRAS DOGs to collect more FIR information for the matched sources. 
Consequently, we selected $59 + 50 = 109$ DOGs (hereafter WISE--SDSS--IRAS/AKARI DOGs) in this work.
The main difference of the WISE--SDSS--IRAS/AKARI DOGs and classical DOGs discovered by \cite{Dey} is the MIR flux; the typical (median) flux density at 22 $\micron$ of our DOG sample is 10.4 mJy that is much brighter than 0.3 mJy at 24 $\micron$ selected by \cite{Dey} (see also Section \ref{D_M_SFR}).

\begin{figure*}
   \centering
   \includegraphics[width=0.95\textwidth]{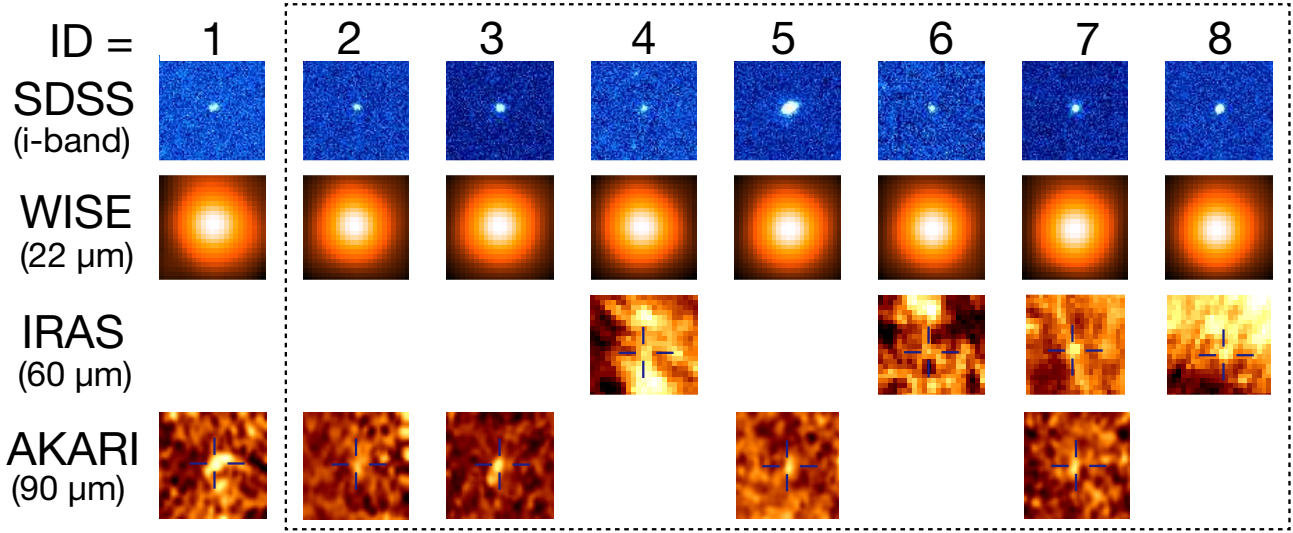}
   \caption{Cutout images at $i$-band (top), 22 $\micron$ (upper-middle), 60 $\micron$ (lower-middle), 100 $\micron$ (bottom) taken by SDSS, WISE, {\it IRAS}, and {\it AKARI}, respectively. Each image is $30\arcsec \times 30\arcsec$ for SDSS and WISE, $30\arcmin \times 30\arcmin$ for {\it IRAS}, and $10\arcmin \times 10\arcmin$ for {\it AKARI} in size, respectively. Not that we obtained the stellar mass and SFR for 7 objects with ID=2-8 (see Section \ref{result_SED}).}
   \label{cutout}
   \end{figure*}

For the 109 WISE-SDSS-IRAS/AKARI DOGs, we compiled spectroscopic redshift information by utilizing the NASA/IPAC Extragalactic Database (NED\footnote{http://ned.ipac.caltech.edu/}) and the Set of Identifications, Measurements, and Bibliography for Astronomical Data (SIMBAD) database\footnote{http://simbad.u-strasbg.fr/simbad/}.
For spectroscopically confirmed DOGs, we rejected nearby galaxies with redshift smaller than 0.07 to ensure the reliable photometry at each band because the photometry we employed is not optimized for extended sources (see Section \ref{SED_fitting}).
Finally, we visually checked optical, MIR, and FIR images and excluded a suspicious object that is affected by nearby bright star and/or that cannot be deblended by the SDSS pipeline.
As a result, we selected 8 objects with spectroscopic redshift (hereafter WISE--SDSS--IRAS/AKARI specz DOGs) with $0.07 < z < 1.0$ (its mean redshift is $\sim$ 0.54).
We note that only one object has a SDSS spectrum with meaningful quality because the mean $i$-band magnitude of our DOG sample is $\sim$ 19.4 that is fainter than that of typical SDSS spectroscopic galaxies.
Figure \ref{cutout} shows cutout images at optical, MIR, and FIR wavelength.
All the information of our DOGs sample is tabulated in Table \ref{table}.

\subsection{SED Fitting to derive the IR luminosity}
\label{SED_fitting}
We performed the SED fitting for the 8 WISE--SDSS--IRAS/AKARI specz DOG sample to derive the total IR luminosity, $L_{\mathrm{IR}}$ (8--1000 $\mu$m).
We employed the fitting code SEd Analysis using BAyesian Statistics ({\tt SEABASs}\footnote{http://xraygroup.astro.noa.gr/SEABASs/}: \citealt{Rovilos}) that provides up to three-component fitting (AGN, SF, and stellar component) based on the maximum likelihood method.
For the AGN templates, we utilized the library of \cite{Silva}, which contains torus templates with varying extinction ranging from $N_{\rm H}$ = 0 to $N_{\rm H}$ = 10$^{25}$ cm$^{-2}$.
For the SF templates, we utilized the library of \cite{Chary} and \cite{Mullaney}.
We also used the library of \cite{Polletta} representing optically selected AGNs and SF galaxies (see \citealt{Polletta} for more detail).
For the stellar templates, {\tt SEABASs} gives a library of 1500 synthetic stellar templates from \cite{Bruzual} stellar population models with solar metallicity and a range of SF histories and ages assuming a \cite{Chabrier} initial mass function (IMF), and each model are reddened using a \cite{Calzetti} dust extinction law.
In order to derive the total IR luminosity and stellar mass with small uncertainties, we used data only with {\tt fqual\_25/60/100} $\geq$ 2 and {\tt fqual\_65/69/140/160} = 3 for {\it IRAS} and {\it AKARI} data, respectively.
Eventually, we performed the SED fitting for 10--13 photometric points among $u$, $g$, $r$, $i$, $z$, $J$, $H$, $K$s-band and 3.4, 4.6, 12, 22, 25, 60, 65, 90, 100, 140, and 160 $\micron$ with SDSS, the TwoMicron All Sky Survey \citep[2MASS: ][] {Skrutskie} Point Source Catalog \citep[PSC: ][]{Cutri_03}, {\it WISE}, {\it IRAS} and {\it AKARI} data, and estimated the total IR luminosity and stellar mass.

We employed the profile-fit magnitude and CmodelMag for each source in the {\it WISE} and SDSS catalogs, respectively, which traces total flux (hereinafter we used $u$, $g$, $r$, $i$, and $z$ as a shorthand alias for CmodelMag).
For the remaining photometry in 2MASS, {\it IRAS}, and {\it AKARI}, we used the default magnitude or flux in each catalog.
Note that profile-fit photometry in {\it WISE} and default magnitude in 2MASS PSC are optimized for point sources and may underestimate the true brightness of extended sources.
However, since we selected point like sources by adopting flags and redshift cut (see Section \ref{Selection}) and visually confirmed it, the photometry we used is reliable.

We took into account the equilibrium between the energy absorbed from the stellar component and the energy emitted in the IR by the SF.
The uncertainties of the derived $L_{\rm IR}$ contains not only statistical error but also systematic error.
{\tt SEABASs} can calculate $L_{\rm IR}$ for ``every'' trial fit and estimate the likelihood value (corresponding to the chi-square) for each case, and provides us the uncertainties as the 2$\sigma$ confidence interval.
Therefore the influence of the difference between the inputed SED templates on the derived $L_{\rm IR}$ is included in the uncertainty.

\floattable
\begin{splitdeluxetable*}{ccrllcccrrrBcccrrrrrrBrrrrrrrrr} 
\tabletypesize{\small}
\rotate
\tablewidth{0pt} 
\tablecaption{The 8 WISE--SDSS--IRAS/AKARI specz DOGs identified in this work. \label{table}}
\tablecolumns{29}
\tablenum{2}
\tablewidth{0pt}
\tablehead{
\colhead{ID} &
\colhead{FIS ID} &
\colhead{objname} &
\colhead{R.A.\tablenotemark{a}} &
\colhead{Decl.\tablenotemark{a}} & 
\colhead{redshift\tablenotemark{b}} & 
\colhead{$u$mag} &
\colhead{$g$mag} &
\colhead{$r$mag} &
\colhead{$i$mag} &
\colhead{$z$mag} &
\colhead{$j$mag} &
\colhead{$h$mag} &
\colhead{$ks$mag} &
\colhead{3.4 $\micron$ flux } &
\colhead{4.6 $\micron$ flux } &
\colhead{12 $\micron$ flux } &
\colhead{22 $\micron$ flux } &
\colhead{25 $\micron$ flux (fqual\_25)} &
\colhead{60 $\micron$ flux (fqual\_60)} &
\colhead{65 $\micron$ flux (fqual\_65)} &
\colhead{90 $\micron$ flux (fqual\_90)} &
\colhead{100 $\micron$ flux (fqual\_100)} &
\colhead{140 $\micron$ flux (fqual\_140)} &
\colhead{160 $\micron$ flux (fqual\_160)} &
\colhead{$\log \,L_{\rm IR}$} &
\colhead{$\log\,M_{*}$} &
\colhead{$\log$ SFR}
\\
\colhead{} &
\colhead{} &
\colhead{} &
\colhead{hms} &
\colhead{dms} & 
\colhead{} &
\colhead{AB mag} &
\colhead{AB mag} &
\colhead{AB mag} &
\colhead{AB mag} &
\colhead{AB mag} &
\colhead{AB mag} &
\colhead{AB mag} &
\colhead{AB mag} &
\colhead{mJy} &
\colhead{mJy} &
\colhead{mJy} &
\colhead{mJy} &
\colhead{mJy} &
\colhead{mJy} &
\colhead{mJy} &
\colhead{mJy} &
\colhead{mJy} &
\colhead{mJy} &
\colhead{mJy} &
\colhead{$L_{\sun}$} &
\colhead{$M_{\sun}$} &
\colhead{$M_{\sun}$/yr}
}
\startdata
 1 & 5007582 & AKARI J00260+1041 & 00:26:06.6 & +10:41:26.5 & 0.57 & 21.89 $\pm$ 0.27 & 21.37 $\pm$ 0.06 & 20.37 $\pm$ 0.04 & 19.92 $\pm$ 0.05 & 19.79 $\pm$ 0.13 & --- & --- & --- & 0.15 $\pm$ 0.01 & 0.13 $\pm$ 0.01 & 7.09 $\pm$ 0.29 & 29.68 $\pm$ 1.57 & --- & --- & 359.96 $\pm$ 250.89 (1) & 417.40 $\pm$ 50.29 (3) & --- & 407.51 $\pm$ 231.02 (1) & --- & --- & ---  & ---  \\
 2 & 5204882 & AKARI J09150+2418 & 09:15:01.7 & +24:18:12.1 & 0.84 & 20.91 $\pm$ 0.06 & 20.59 $\pm$ 0.03 & 20.28 $\pm$ 0.03 & 19.95 $\pm$ 0.03 & 19.01 $\pm$ 0.04 & 17.43 $\pm$ 0.17 & 16.59 $\pm$ 0.12 & 15.65 $\pm$ 0.04 & 8.20 $\pm$ 0.17 & 16.61 $\pm$ 0.31 & 41.92 $\pm$ 0.63 & 96.12 $\pm$ 2.41 & --- & --- & --- & 396.88 $\pm$ 64.55 (3) & --- & 40.05 $\pm$ 258.74 (1) & --- & 13.52$_{-0.01}^{+0.04}$ & 12.60$_{-0.18}^{+0.18}$ & 2.56$_{-0.19}^{+0.25}$ \\
 3 & 5291698 & AKARI J13070+2338 & 13:07:00.6 & +23:38:05.1 & 0.28 & 21.46 $\pm$ 0.18 & 20.58 $\pm$ 0.03 & 19.38 $\pm$ 0.02 & 19.02 $\pm$ 0.02 & 18.56 $\pm$ 0.05 & 17.66 $\pm$ 0.16 & 16.44 $\pm$ 0.07 & 15.31 $\pm$ 0.04 & 8.65 $\pm$ 0.18 & 14.71 $\pm$ 0.28 & 26.38 $\pm$ 0.42 & 75.79 $\pm$ 2.14 & --- & --- & 453.78 $\pm$ 255.58 (1) & 557.62 $\pm$ 61.57 (3) & --- & 507.33 $\pm$ 244.22 (1) & 53.72 $\pm$ 338.50 (1) & 12.33$_{-0.02}^{+0.06}$ & 9.86$_{-0.13}^{+0.10}$ & 1.95$_{-0.04}^{+0.12}$ \\
 4 &--- &  IRAS F13073+6057 & 13:09:16.9 & +60:42:08.9 & 0.64 & 22.65 $\pm$ 0.32 & 22.91 $\pm$ 0.15 & 21.59 $\pm$ 0.12 & 20.47 $\pm$ 0.06 & 19.42 $\pm$ 0.11 & 18.76 & 17.50 $\pm$ 0.24 & 16.64 $\pm$ 0.11 & 2.23 $\pm$ 0.05 & 4.23 $\pm$ 0.08 & 10.86 $\pm$ 0.22 & 26.94 $\pm$ 1.07 & 84.43 $\pm$ 17.73 (1) & 194.70 $\pm$ 42.83 (3) &--- &--- & 818.30 $\pm$ 171.84 (1) &--- &--- & 12.75$_{-0.11}^{+0.07}$ & 11.67$_{-0.11}^{+0.27}$ & 2.35$_{-0.53}^{+0.16}$ \\
 5 & 5317429 & AKARI J14063+0103 & 14:06:38.2 & +01:02:54.5 & 0.24 & 19.40 $\pm$ 0.05 & 18.69 $\pm$ 0.01 & 18.10 $\pm$ 0.01 & 17.70 $\pm$ 0.01 & 17.64 $\pm$ 0.03 & 17.91 $\pm$ 0.20 & 17.51 $\pm$ 0.21 & 16.65 $\pm$ 0.13 & 3.09 $\pm$ 0.07 & 8.87 $\pm$ 0.18 & 75.79 $\pm$ 1.07 & 236.16 $\pm$ 4.45 & --- & --- & 598.65 $\pm$ 252.74 (1) & 684.26 $\pm$ 55.28 (3) & --- & 1332.88 $\pm$ 255.74 (1) & --- & 12.63$_{-0.07}^{+0.06}$ & 10.28$_{-0.17}^{+0.01}$ & 2.59$_{-0.10}^{+0.06}$ \\
 6 &--- &  IRAS F14481+4454 & 14:49:53.6 & +44:41:50.3 & 0.67 & 21.16 $\pm$ 0.08 & 20.74 $\pm$ 0.03 & 20.56 $\pm$ 0.03 & 20.47 $\pm$ 0.04 & 20.02 $\pm$ 0.14 & --- & --- & --- & 2.02 $\pm$ 0.04 & 5.50 $\pm$ 0.10 & 24.57 $\pm$ 0.41 & 65.77 $\pm$ 1.65 & 85.09 $\pm$ 17.87 (2) & 189.70 $\pm$ 32.25 (3) &--- &--- & 500.50 $\pm$ 115.12 (1) &--- &--- & 13.29$_{-0.03}^{+0.12}$ & 9.16$_{-0.17}^{+0.08}$ & 3.16$_{-0.05}^{+0.15}$ \\
 7 & 5365228 & AKARI J15324+3242 & 15:32:44.0 & +32:42:46.6 & 0.93 & 20.18 $\pm$ 0.05 & 19.91 $\pm$ 0.02 & 19.53 $\pm$ 0.02 & 19.23 $\pm$ 0.02 & 18.40 $\pm$ 0.04 & --- & --- & --- & 0.41 $\pm$ 0.01 & 0.82 $\pm$ 0.02 & 10.55 $\pm$ 0.23 & 48.31 $\pm$ 1.22 & 71.03 $\pm$ 21.31 (2) & 234.10 $\pm$ 35.12 (3) & 61.96 $\pm$ 191.01 (1) & 398.36 $\pm$ 46.21 (3) & 711.80 $\pm$ 170.83 (1) & 983.85 $\pm$ 189.00 (1) & 433.56 $\pm$ 325.80 (1) & 13.43$_{-0.01}^{+0.04}$ & 10.85$_{-0.00}^{+0.02}$ & 2.88$_{-0.04}^{+0.13}$ \\
 8 &--- &  IRAS F23497-0448 & 23:52:15.1 & -04:32:10.5 & 0.16 & 19.98 $\pm$ 0.07 & 19.47 $\pm$ 0.01 & 18.86 $\pm$ 0.01 & 18.74 $\pm$ 0.02 & 19.19 $\pm$ 0.09 & --- & --- & --- & 0.19 $\pm$ 0.01 & 1.25 $\pm$ 0.03 & 18.78 $\pm$ 0.39 & 128.23 $\pm$ 2.98 & 311.50 $\pm$ 93.45 (1) & 396.30 $\pm$ 67.37 (3) &--- &--- & 571.50 $\pm$ 165.73 (1) &--- &--- & 11.98$_{-0.07}^{+0.09}$ & 9.65$_{-0.01}^{+0.01}$ & 1.94$_{-0.07}^{+0.09}$ \\
\enddata
\tablenotetext{a}{The coordinates in the SDSS DR12.}
\tablenotetext{b}{NED.}
\end{splitdeluxetable*}

\section{Results}
  
\subsection{Result of SED fitting}
\label{result_SED}
   \begin{figure*}
   \centering
   \includegraphics[width=0.95\textwidth]{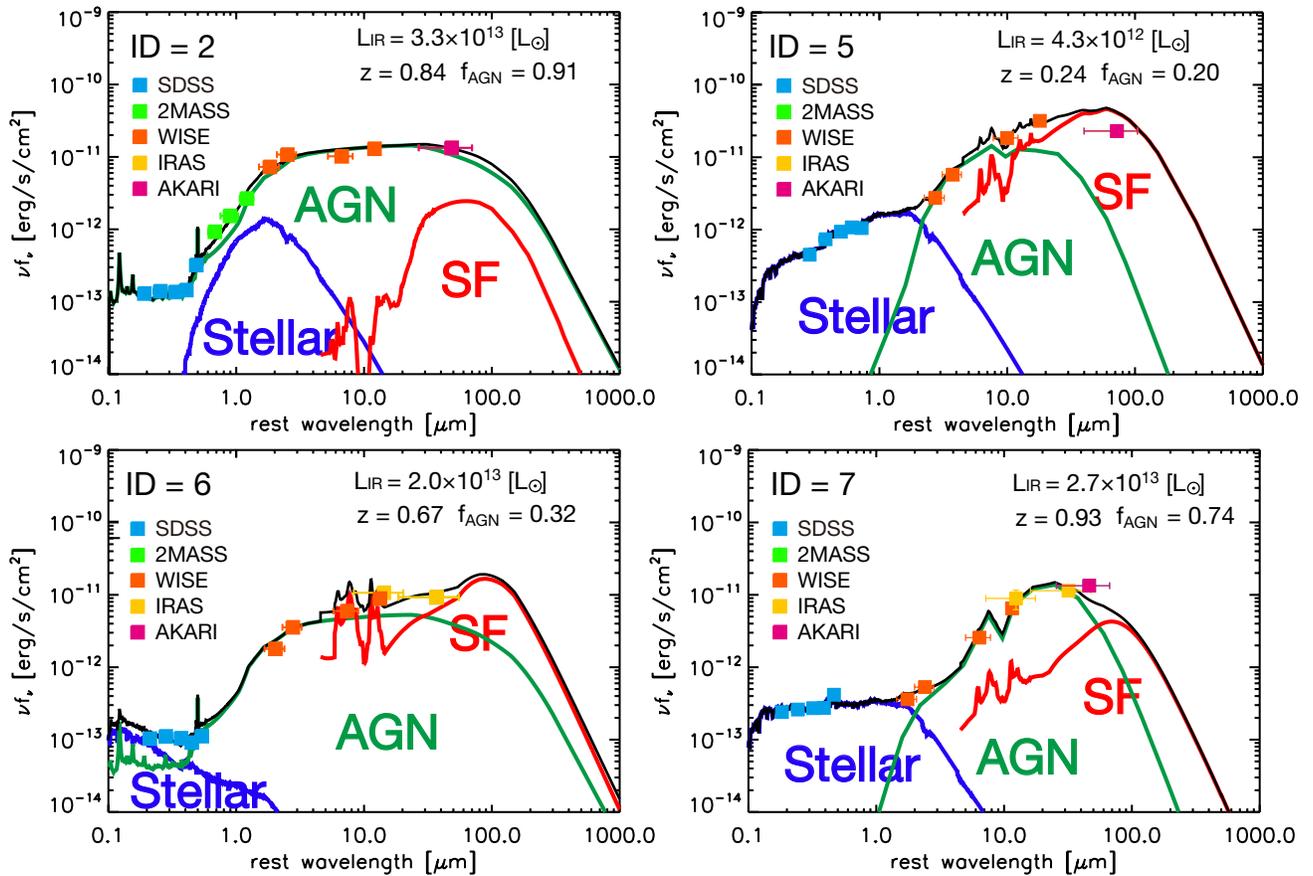}
   \caption{Examples of the SED fitting for our DOG sample. The blue, green, orange, yellow, and pink square represent the data from the SDSS, 2MASS, {\it WISE}, {\it IRAS}, and {\it AKARI}, respectively.The contribution from the stellar, AGN, SF components to the total SEDs are shown in blue, green, and red lines, respectively. The black solid line represents the resultant (the combination of the stellar, AGN, and SF) SEDs. We only used data with {\tt fqual\_25/60/100} $\geq$ 2 and {\tt fqual\_65/69/140/160} = 3 for {\it IRAS} and {\it AKARI} data, respectively, for the SED fitting.}
   \label{SEABAS}
   \end{figure*}
Figure \ref{SEABAS} shows the example of the SED fitting.
The best-fit AGN template for our IR-bright DOGs sample tends to favor the ``torus'' template presented by \cite{Silva} or \cite{Polletta}, which is consistent with the report by \cite{Tsai} based on the WISE-selected IR luminous sources.
A remarkable aspect we found is that IR-bright DOGs have a flat SED in MIR region which provides us a clue of an empirical relation of their MIR and IR luminosities (see Section \ref{LL}).
It should be noted that one object with ID=1 cannot be well-fitted by {\tt SEABASs} code.
Its best-fit SED in FIR region provided significantly lower than observed flux at 90 $\micron$.
One possible reason is that this object has only one photometric point in the FIR regime that could not be enough to constrain the FIR SEDs.
Another possibility is that its 90 $\mu$m flux density might be overestimated due to the deblending issue and thus SEABASs with considering energy balance between UV/optical and IR cannot reproduce the FIR emission. 
Therefore, we excluded this object for the SED fitting and we derive the stellar mass, total IR luminosity, and SFR for the remaining 7 DOGs with ID=2-8 (see Table \ref{table}).
Hereinafter, we focus on these objects.

\subsection{Energy contribution of AGN to the IR luminosity}

\begin{figure}
   \centering
   \includegraphics[width=0.45\textwidth]{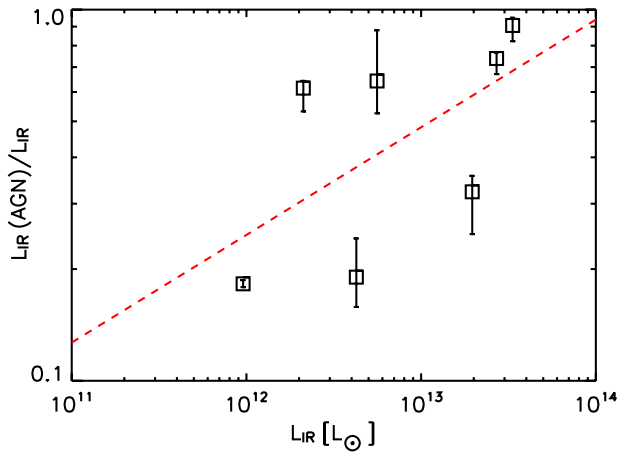}
   \caption{The luminosity contribution of AGN to the IR luminosity, $L_{\rm IR}$ (AGN)$ / L_{\rm IR}$, as a function of IR luminosity. The red dotted line shows the best-fit linear function.}
   \label{LAGN}
   \end{figure}

Since {\tt SEABASs} executes the three-component SED fitting of stellar, AGN, and SF, we can calculate the energy contribution of each component to the IR luminosity.
Figure \ref{LAGN} shows the luminosity contribution of AGN to the IR luminosity ($L_{\rm IR}$ (AGN)$ / L_{\rm IR}$) as a function of IR luminosity.   
We found that the energy contribution of AGN to the IR luminosity increases with increasing the IR luminosity.
This result is in good agreement with those from {\it AKARI} selected LIRGs/ULIRGs \citep[e.g., ][]{Lee_12,Ichikawa} and those from IR-faint DOGs \citep[e.g., ][]{Riguccini},
in the sense that more IR-luminous sources tend to be more AGN-dominated. 
The fact that the luminous IR sources tend to be more AGN dominated relatively reported by several authors can be applicable for IR-bright DOGs.

\section{Discussions}

\subsection{Predicting $L_{\rm IR}$ from 22 and 90 $\micron$ flux density}
\label{LL}
Here we discuss the correlation among the MIR luminosity, FIR luminosity, and total IR luminosity for IR-bright DOGs.
As shown in Figure \ref{SEABAS}, the SEDs of DOGs in MIR regime appears flat, which gives the 
possibility to estimate their IR luminosities from a MIR luminosity at ``observed-frame'' without considering $k$-correction.
We derive 22 and 90 $\mu$m luminosity density at observed frame, $L^{\rm obs}_{\nu}$ (22 $\micron$ or 90 $\micron$), just from the observed flux density by multiplying $4\pi d_{\rm L}^2$ for each DOG, where $d_{\rm L}$ is luminosity distance.

   \begin{figure*}
   \centering
   \includegraphics[width=\textwidth]{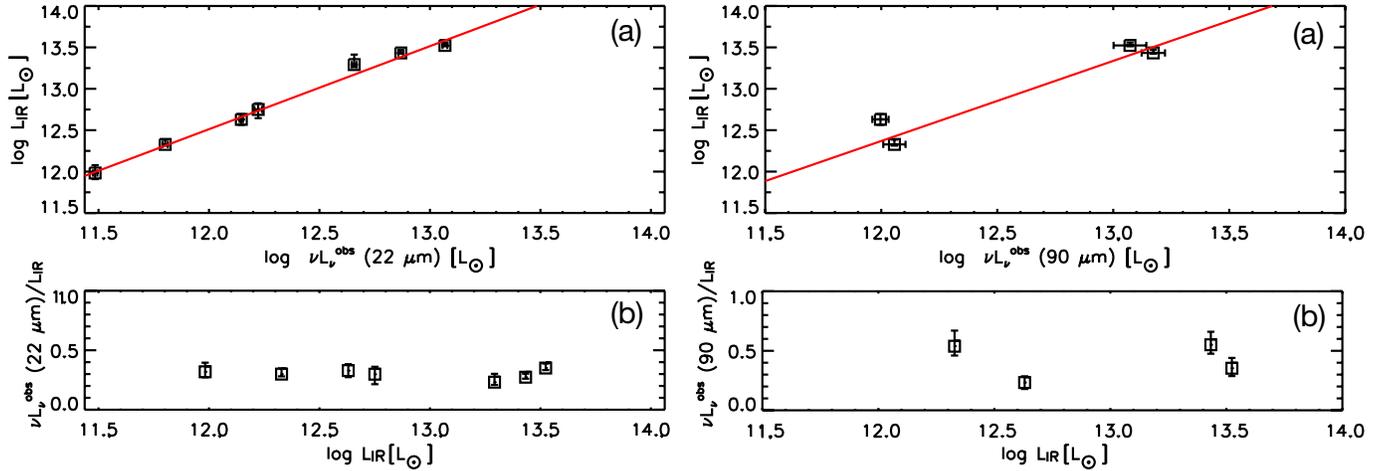}
   \caption{(a) {\it WISE} 22 $\micron$ luminosity at observed frame vs. IR luminosity (left) and {\it AKARI} 90 $\micron$ luminosity at observed frame vs. IR luminosities (right). The red dotted line shows the best-fit linear function. (b) The ratio of 22 $\micron$ (left) and 90 $\micron$ (right), and IR luminosity as a function of IR luminosity.}
   \label{LIR}
   \end{figure*}
Figure \ref{LIR}(a) shows the relation between the 22 and 90 $\micron$ luminosity in the observed frame and IR luminosity.
We see tight correlations between 22 $\micron$ luminosity and IR luminosity, and 90 $\micron$ luminosity and IR luminosity.
Figure \ref{LIR}(b) shows the ratio of the 22 and 90 $\micron$, and IR luminosity as a function of IR luminosity.
The ratio of {\it WISE} 22 $\micron$ and IR luminosity have similar value regardless of IR luminosity, which suggests that 22 $\micron$ luminosity is a more linear relationship with IR luminosity.  
We obtained the following conversion formulae:
\begin{eqnarray}
\log L_{\rm IR} & = & (1.00 \pm 0.02) \log \,[\nu L^{\rm obs}_{\nu} \,(22\, \micron)] \nonumber \\
				& + & (0.48 \pm 0.28), \\
\log L_{\rm IR} & = & (0.97 \pm 0.06) \log \,[\nu L^{\rm obs}_{\nu} \,(90\, \micron)] \nonumber \\
				& + & (0.76 \pm 0.79). 
\end{eqnarray}
The Spearman rank correlation coefficients for each relationship are $\sim$ 1.00 and 0.60 with null hypothesis probability $P \sim 0$ and 4.0 $\times 10^{-1}$, respectively, indicating that 22 $\micron$ luminosity can be used to predict the total IR luminosity for IR-bright DOGs with $0.07 < z < 1.0$ without considering $k$-correction. 
At the same time, we should keep in mind that whether or not this empirical relation is applicable to other galaxies is unknown, and thus this relation may be useful only for IR-bright DOGs.

\subsection{Stellar mass and SFR}
\label{D_M_SFR}
Since {\tt SEABASs} has the advantage of being able to decompose the total SED into stellar, AGN, and SF components, we used IR luminosity contributed from SF activity and convert it to SFR using \cite{Kennicutt}  
equation with the the Chabrier IMF calibrated by \cite{Salim},
\begin{equation}
SFR = \log L_{\rm IR} \, ({\rm SF}) - 9.966.
\end{equation}
where SFR and $L_{\rm IR}$ (SF) are given in unit of $M_{\sun}$ yr$^{-1}$ and $L_{\sun}$, respectively.
For the stellar mass, we used the output from the SED fitting based on {\tt SEABASs} assuming same IMF.
Note that the rest-frame UV continuum may be contributed by scattered light from AGNs particularly for luminous DOGs \cite[e.g., ][]{Hamann}, which induces an uncertainty for estimated stellar mass. 

\subsubsection{One to one comparison with literature}
\label{offset}
   \begin{figure}
   \centering
   \includegraphics[width=0.45\textwidth]{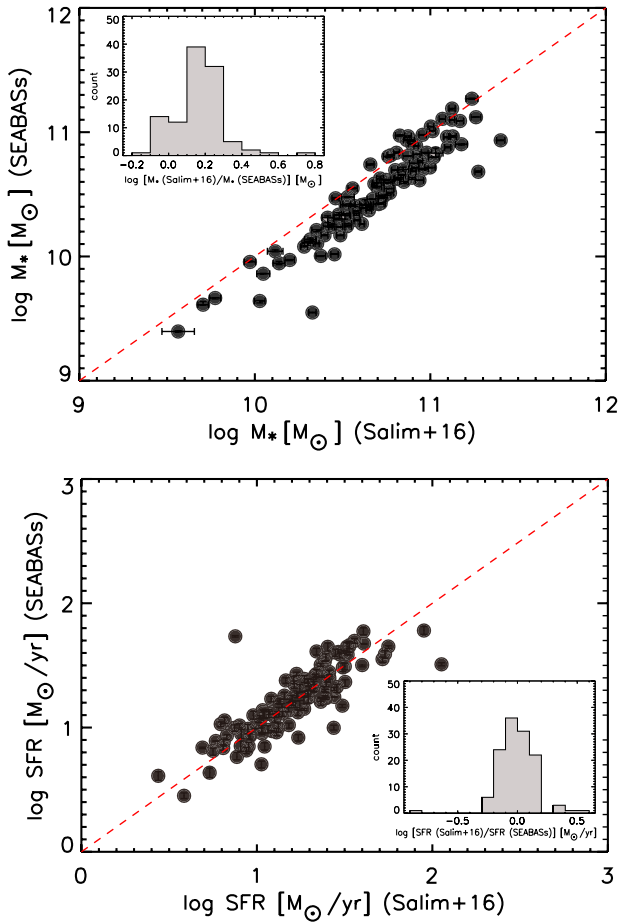}
   \caption{Comparison of stellar mass (top) and SFR (bottom) derived by {\tt SEABASs} we employed in this work and \cite{Salim}. The red dotted line is the one-to-one line. The inserted figure shows the histogram of the ratio of each quantity derived in this work and \cite{Salim} for stellar mass (top) and SFR (bottom), respectively.}
   \label{Comp_S}
   \end{figure}
Before comparing SFR--$M_*$ relation of IR-bright DOG sample with those of other population, we investigate whether or not the estimate of stellar mass and SFR based on {\tt SEABAS} have a systematic offset compared to those derived from previous works using some local galaxies/ULIRGs.
One caution here is that the difference of the assumed IMF also affects the stellar mass and SFR \citep[e.g.,][]{Rieke,Casey}, which induces a systematic offset on SFR--$M_*$ plane.
Therefore, we corrected the stellar mass and SFR in the literature to those assumed Chabrier IMF by dividing them by 1.58 \cite{Salim_07} \citep[see also][]{Tacconi} if needed.

First, we compare stellar mass and SFR with those of local galaxies at $z < 0.3$ estimated by \cite{Salim_07}.
They recently provided a catalog (GALEX--SDSS--WISE Legacy Catalog: GSWLC) of physical properties including stellar masses and SFRs that were derived by the SED fitting following a Bayesian methodology for UV/optical data.
Note that in addition to SFRs derived from the SED fitting, they provided MIR SFRs derived from IR templates based on WISE 22 $\micron$ data to avoid potential systematics.
This is because they do not use the FIR data to derive the SFR that could induce large uncertainties \citep[see e.g.,][]{Toba_16}.
So, we first cross matched the GSWLC catalog with {\it AKARI} FIS BSC ver.2, and derived stellar mass and SFR of matched sources as described above.
We then compared each quantity of them with those in GSWLC where we used MIR-SFR derived from ALLWISE catalog.

Figure \ref{Comp_S} shows the comparison of stellar masses derived from our method and those in GSWLC.
Our estimate of SFR is in good agreement with that in \cite{Salim_07} while the stellar mass we estimated is slightly smaller than that in \cite{Salim_07}; the typical offset of stellar mass is $\sim 0.15$ dex.
This offset is roughly consistent with that results from the comparison between GSWLC and the Max Planck Institute for Astrophysics/Johns Hopkins University (MPA/JHU) catalog \citep{Kauffmann,Brinchmann} \citep[see] [in detail]{Salim_07}.
We should keep in mind this offset when comparing the stellar masses with local SDSS galaxies.

\begin{figure}
   \centering
   \includegraphics[width=0.45\textwidth]{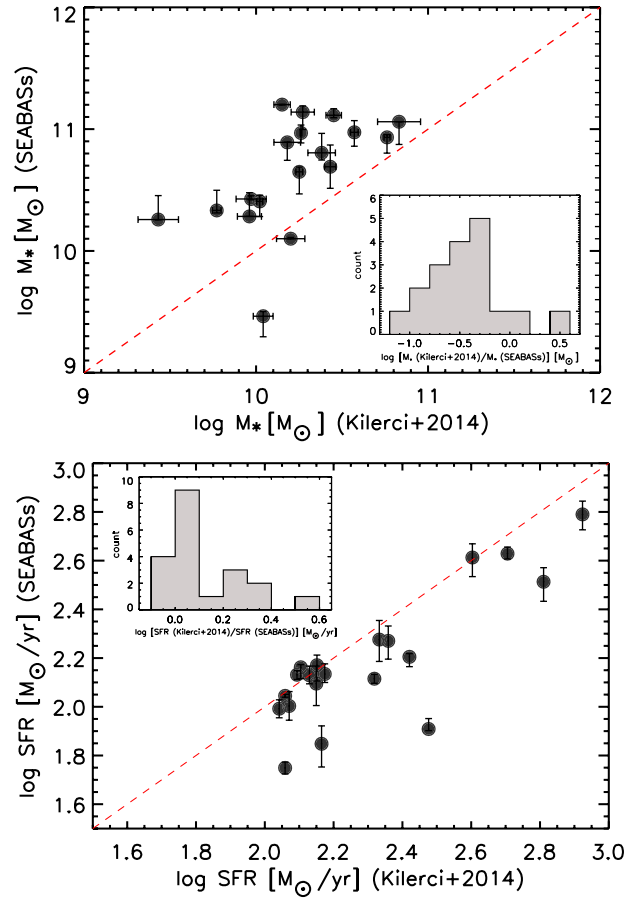}
   \caption{Comparison of stellar mass (top) and SFR (bottom) derived by {\tt SEABASs} we employed in this work and \cite{Kilerci}. The red dotted line is the one-to-one line. The inserted figure shows the histogram of the ratio of each quantity derived in this work and \cite{Kilerci} for stellar mass (top) and SFR (bottom), respectively.}
   \label{Comp_K}
   \end{figure}

Next, we compare stellar mass and SFR with those of local ULIRGs at $z < 0.3$ estimated by \cite{Kilerci}.
They constructed a ULIRG sample by cross-matching the {\it AKARI} FIS BSC version 1 \citep{Yamamura} with the SDSS DR10 \citep{Ahn}.
Figure \ref{Comp_K} shows the comparison of stellar masses and SFRs derived from our method and those in \cite{Kilerci}.
Our estimate of SFR is roughly consistent with that in \cite{Kilerci} although we underestimate significantly SFR for some ULIRGs, while the stellar mass we estimated is obviously larger than that in \cite{Kilerci}; the typical offset of stellar mass is $\sim$ 0.5 dex.
We note that \cite{Kilerci} reported that all of the adopted stellar mass values in their work might be underestimated by $\sim$ 0.5 dex by comparing the derived stellar mass with previous works, which could be a possible interpretation of the discrepancy of our estimate of stellar mass.
We should keep in mind this offset when comparing the stellar mass with local ULIRGs.

\subsubsection{Stellar mass and SFR relation}
\label{D_M_SFR}
We discuss where IR-bright DOGs lie in SFR--$M_*$ plane and compare it with that of the literature.
We first estimated the SFR based on the IR luminosity from SF.
   \begin{figure*}
   \centering
   \includegraphics[width=0.8\textwidth]{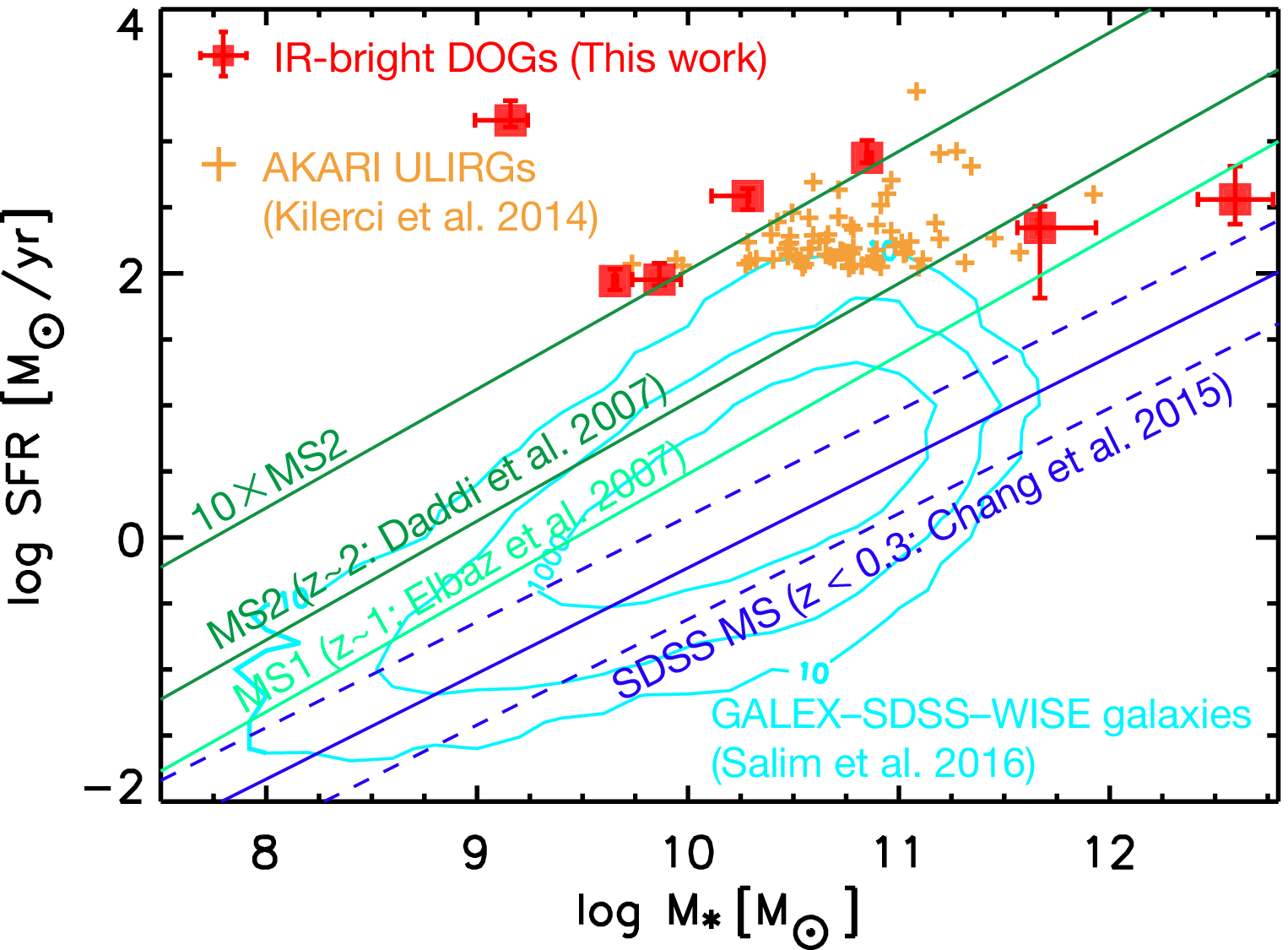}
  \caption{Stellar mass and SFR for 7 IR-bright DOGs (red square) and 75 AKARI-selected ULIRGs \citep{Kilerci}. The blue solid line is main sequence (MS) of normal SF galaxies selected from the SDSS \citep{Chang} with scatter of 0.39 dex (blue dotted line). The cyan contours represent SFR--$M_*$ relation for a sample of GALEX--SDSS--WISE Legacy Catalog \citep[GSWLC: ][]{Salim} at $z < 0.3$. The bin size is 0.2 $\times$ 0.2 in the units given in the plot. The light green line is MS of normal SF galaxies at $z$ = 1 \citep{Elbaz} while the dark green lines are MS of SF galaxies at $z$ = 2 \citep{Daddi} and 10 times above this relationship.}
   \label{M_SFR}
   \end{figure*}
   
Figure \ref{M_SFR} shows the resultant stellar mass and SFR relation for our DOG sample, the main-sequence (MS) sample for star forming galaxies selected by the SDSS and {\it WISE} \citep{Chang}, and selected by the Galaxy Evolution Explorer ({\it GALEX}) satellite \citep{Martin}, SDSS, and {\it WISE} \citep{Salim}.
The stellar mass and SFR of the MS presented by \cite{Elbaz} and \cite{Daddi} for star forming galaxies at $z$ = 1 and 2, respectively, are also shown in Figure \ref{M_SFR}.
Note that we corrected a possible offset of stellar masses discussed in Section \ref{offset} for local SDSS sample provided by \citep{Salim}.
We remind that we corrected the stellar mass and SFR in the literature to those assumed Chabrier IMF if needed.
We found that most IR-bright DOGs lie above these relations significantly although the redshift of our DOG sample is less than 1.0.
They cover a locus of merger-driven starburst galaxies \citep[e.g., ][]{Rodighiero}, indicating that our IR-bright DOG sample detected by {\it IRAS} and/or {\it AKARI} is basically starburst galaxies.
The stellar mass and SFR of a ULIRG sample presented by \cite{Kilerci} are also plotted.
We also corrected a possible offset of stellar masses discussed in Section \ref{offset} for them.
We found that some IR-bright DOGs have larger SFR value given a same stellar mass (although remaining objects overlapped with local ULIRGs in SFR--$M_*$ plane.) 
This is partially due to the difference of the redshift distribution; redshift of sample in \cite{Kilerci} is less than 0.5. 
However, it is difficult to conclude this offset is statistically robust due to small sample size, and thus it will be future work.
The tendency of the large offset of IR-bright DOG from MS in SFR--$M_*$ plane is likely to be inconsistent with that of IR-faint DOGs \citep{Kartaltepe,Riguccini}.
They reported that IR-faint DOGs with no significant AGN contribution are mainly located within the star-forming MS, although some author reported that they are widely distributed on SFR--$M_*$ plane \citep{Calanog,Corral}.
Note that our sample is relatively low-redshift ($z \sim 0.54$) compared with IR-faint DOG sample ($z \sim 2$).
Also, the MIR flux range of them is significantly different.
Taking these results into account, the SF properties of IR-bright DOGs is not necessary to same as IR-faint (i.e., classical) DOGs.
This work enables us to constrain SFR--$M_*$ relation of previously unknown, IR-bright DOGs whose properties differ from those of classical, IR-faint DOGs, for the first time.

It should be noted that SFR--$M_*$ relation of MS is also related to its other physical properties such as 
metallicity \citep[e.g.,][]{Mannucci}, molecular gas fraction \citep[e.g.,][]{Daddi_10,Sargent}, and starburst compactness \citep[e.g.,][]{Elbaz_11}.
However, a full exploitation of how each physical property of IR-bright DOGs affects the SFR--$M_*$ relation requires a further observation to derive each quantity, which is beyond the scope of this paper, and will be a future work.
We here just discuss a possibility of that the large offset of our DOG sample from MS can be explained depending on the metallicity of IR-bright DOGs and MS.
Since the dispersion of SFR and stellar mass due to the metallicity of MS is 0.2-0.4 dex \citep{Savaglio,Lara-Lopez}, the large offset of IR-bright DOGs cannot be explained only by metallicity. 
Therefore, taking the redshift evolution of SFR--$M_*$ relation \citep{Whitaker,Lee_15,Tomczak} and uncertainty of SFR--$M_*$ for MS due to the dispersion of at least metallicity into account, our IR-bright DOGs detected by {\it IRAS} and/or {\it AKARI} seems to be more specific population compared with IR-faint DOGs regarding the SFR--$M_*$ relation.

\acknowledgments
The authors appreciate the referee's thoughtful feedback which improved the manuscript.
This research is based on observations with AKARI, a JAXA project with the participation of ESA.
This research has made use of the NASA/ IPAC Infrared Science Archive, which is operated by the Jet Propulsion Laboratory, California Institute of Technology, under contract with the National Aeronautics and Space Administration.
This publication makes use of data products from the Wide-field Infrared Survey Explorer, which is a joint project of the University of California, Los Angeles, and the Jet Propulsion Laboratory/California Institute of Technology, funded by the National Aeronautics and Space Administration. 
Funding for SDSS-III has been provided by the Alfred P. Sloan Foundation, the Participating Institutions, the National Science Foundation, and the U.S. Department of Energy Office of Science. 
The SDSS-III web site is http://www.sdss3.org/.
SDSS-III is managed by the Astrophysical Research Consortium for the Participating Institutions of the SDSS-III Collaboration including the University of Arizona, the Brazilian Participation Group, Brookhaven National Laboratory, Carnegie Mellon University, University of Florida, the French Participation Group, the German Participation Group, Harvard University, the Instituto de Astrofisica de Canarias, the Michigan State/Notre Dame/JINA Participation Group, Johns Hopkins University, Lawrence Berkeley National Laboratory, Max Planck Institute for Astrophysics, Max Planck Institute for Extraterrestrial Physics, New Mexico State University, New York University, Ohio State University, Pennsylvania State University, University of Portsmouth, Princeton University, the Spanish Participation Group, University of Tokyo, University of Utah, Vanderbilt University, University of Virginia, University of Washington, and Yale University.
This publication makes use of data products from the Two Micron All Sky Survey, which is a joint project of the University of Massachusetts and the Infrared Processing and Analysis Center/California Institute of Technology, funded by the National Aeronautics and Space Administration and the National Science Foundation.
This publication makes use of data products from the Two Micron All Sky Survey, which is a joint project of the University of Massachusetts and the Infrared Processing and Analysis Center/California Institute of Technology, funded by the National Aeronautics and Space Administration and the National Science Foundation.
This research has made use of the NASA/ IPAC Infrared Science Archive, which is operated by the Jet Propulsion Laboratory, California Institute of Technology, under contract with the National Aeronautics and Space Administration.
Y.Toba and W.-H.Wang acknowledge the support from the Ministry of Science and Technology
of Taiwan (MOST 102-2119-M-001-007-MY3 and 105-2112-M-001-029-MY3).
TN is financially supported by the Japan Society for the Promotion of Science (JSPS) KAKENHI (16H01101 and 16H03958).

\end{document}